\documentclass{aa}
\usepackage{graphicx}

\usepackage{txfonts}
\begin{document}
   \title{Local Group timing in Milgromian dynamics}
   \subtitle{A past Milky Way-Andromeda encounter at $z>0.8$}
   \titlerunning{Local Group timing in MOND}

   \author{H. Zhao
          \inst{1}
          \and
          B. Famaey
          \inst{2}
          \and
          F. L\"ughausen
         \inst{3}
          \and
          P. Kroupa
          \inst{3}
          }

   \institute{SUPA, School of Physics and Astronomy, University of St Andrews, UK\\
              \email{hz4@st-andrews.ac.uk}
         \and
         Observatoire Astronomique,  Universit\'e de Strasbourg, CNRS UMR 7550, Strasbourg, France
\\ \email{benoit.famaey@astro.unistra.fr}
         \and
         AIfA, Universit\"at Bonn, Germany}

   \date{Received ; accepted}

 
  \abstract
   {The Local Group timing has been one of the first historical probes of the missing mass problem. Whilst modern cosmological probes indicate that pure baryonic dynamics is not sufficient on the largest scales, nearby galaxies and small galaxy groups persistently obey Milgrom's MOND law, which implies that dynamics at small scales is possibly entirely predicted by the baryons.}
   {Here, we investigate the Local Group timing in this context of Milgromian dynamics.}
   {Making use of the latest measured proper motions and radial velocities for Andromeda and the Magellanic clouds, we integrate backwards their orbits by making use of the Milgromian two-body equation of motion.}
   {We find that, with the currently measured proper motions and radial velocity of M31, MOND would imply that the Milky Way and M31 first moved apart via Hubble expansion after birth, but then necessarily got attracted again by the Milgromian gravitational attraction, and had a past fly-by encounter before coming to their present positions. This encounter would most probably have happened 7 to 11 Gyr ago ($0.8 < z < 3$). The absence of a dark matter halo and its associated dynamical friction is necessary for such a close encounter not to trigger a merger. Observational arguments which could exclude or favour such a past encounter would thus be very important in view of falsifying or vindicating Milgromian dynamics on the scale of the Local Group. Interestingly, the closest approach of the encounter is small enough ($<55$~kpc) to have severe consequences on the disk dynamics, including perhaps thick disk formation, and on the satellite systems of both galaxies. Integrating back the orbits of the Magellanic clouds, we find that, for the nominal values of their proper motions, they were close to pericenter at the time of the encounter, suggesting that their dynamics and/or origin might possibly be related to the event. The ages of the satellite galaxies and of the young halo globular clusters, all of which form the vast polar structure around the Milky Way, are consistent with these objects having been born in this encounter.}
  {}
   \maketitle
%
\section{Introduction}

The Local Group (LG) timing (e.g., Lynden-Bell 1981, Peebles 1989) has been one of the historical probes of the missing mass problem. 
At the end of the fifties, Kahn \& Woltjer (1959) noted that the Milky Way (MW) and Andromeda (M31) galaxies were approaching each other, thus overcoming cosmic expansion. They concluded that, by assuming that they were initially formed close together, the MW-M31 system had to be at least 20 times more massive than the stellar mass to actually overcome cosmic expansion and reach its current velocity and position. In its simplest model-version, the LG consists of the MW and M31 as two isolated point masses that moved apart due to the Hubble expansion, then slowed down and moved towards each other again, which could not possibly happen in Newtonian gravity without dark matter (DM).   

Since then, numerous other cross-matching pieces of evidence, culminating with the latest data release of the Planck mission (Ade et al. 2013), have accumulated to indicate the apparent need for missing non-baryonic fields (behaving as a dissipationless dust fluid) on the largest scales of the Universe. However, this does not necessarily and univoquely imply the existence of stable neutral DM particles at galactic scales. The observed phenomenology in a wide range of near-field galaxy data (e.g., Famaey \& McGaugh 2012) indeed indicates that the gravitational field can entirely be predicted by the baryons, independently from the history and environment of each galaxy. This is at odds with the a priori predictions from collisionless DM particles since it would then imply a large amount of fine-tuning in the baryons-feedback for a large sample of independent systems. This adds up to the numerous other small-scale problems reviewed in, e.g., Kroupa et al. (2010). 

An alternative is to consider that dynamics on small scales is {\it fundamentally} governed by the empirical MOND law of Milgrom (1983), which summarizes the above phenomenology by stating that for gravitational accelerations below $a_0 \simeq 10^{-10} {\rm m} \, {\rm s}^{-2}$, the effective gravitational attraction $g$ approaches $(g_N a_0)^{1/2}$ where $g_N \propto 1/r^2$ is the usual Newtonian gravitational acceleration calculated from the observed distribution of baryonic matter. This might be reconciled with cosmological data through covariant theories involving, e.g., massive vector fields (e.g., Zhao \& Li 2010), bimetric theories with twin matter fields (e.g., Milgrom 2009), the presence of a dipolar dark fluid (e.g., Blanchet \& Le Tiec 2009), etc. These fields are fundamentally different from cold DM particles on galaxy scales: for instance, they would not induce dynamical friction in galaxies between the stars and DM particles, contrary to what happens in the standard context, and in the weak-field limit, all these theories do lead to the MOND phenomenology. To reconcile these theories with galaxy clusters, one might resort to the non-trivial effect of these new fields on large scales (e.g., Dai et al. 2008), or to additional particles that would {\it not} cluster on galaxy scales (e.g., Angus et al. 2007). On the other hand, on the scale of galaxy groups, data seem to be in agreement with the MOND phenomenology without any addition (Milgrom 2002). More generally, weak lensing on the scales of hundreds of kpc indicates that the effective gravity falls off as $1/r$ for luminosities comparable to those of the LG (Brimioulle et al. 2013, Milgrom 2013). We thus investigate here what this phenomenology would imply for the history of the LG.

Motivated by the recent Hubble Space Telescope (HST) measurement of the proper motion of M31 (Sohn et al. 2012), which has considerably reduced the error bars, we investigate the timing argument in the MOND context. Precise measurements of the proper motion of the Magellanic clouds have been obtained too (Kallivayalil et al. 2013). These HST measurements have led to a few recent studies (i) estimating the dynamical mass of the LG in the standard context (van der Marel et al. 2012a) and showing that the timing argument led to an increase in dynamical mass of about 10\% w.r.t. other estimates, (ii) studying the future encounter of the MW and M31 (van der Marel et al. 2012b), or (iii) studying the variety of solutions allowed for the history and dynamics of the LG (Peebles \& Tully 2013). 

In the context of MOND, the situation is slightly different, as there is not as much freedom on the baryonic masses of galaxies as there is on their putative DM mass in the standard context. Once we know the baryonic masses, distances and velocities of galaxies, as well as the external field from Large Scale Structure, a unique history can be traced backwards. Making use of the Milgromian two-body equation of motion, we thus investigate what these recent measurements would imply in the context of MOND. In Sect.~2, we derive the equation of motion in the context of Hubble expansion, we then intergrate backwards the orbits of some LG galaxies in Sect.~3, and discuss the results in Sect.~4.

\section{Two-body equation of motion in a cosmological context}

Even in their classical non-covariant form, MOND theories (Bekenstein \& Milgrom 1984; Milgrom 2010) are inherently non-linear and require to solve complicated modified Poisson field equations (e.g., Tiret \& Combes 2007; L\"ughausen et al. 2013). But in a simple two-body configuration of two point-like masses $m_1$ and $m_2$ separated by a distance $r_{12}$, they boil down to a two-body force of the form (see, e.g., Milgrom 1994; Zhao \& Famaey 2010; Zhao et al. 2010) :
\begin{equation}
F_{\rm 12} 
= \frac{G m_1 m_2}{r_{12}^2} [1 + y^{-1/2}], 
~ y \equiv \left[\frac{\sqrt{G(m_1+m_2) a_0}}{r_{12} Q a_0}\right]^2,
\end{equation}
where $Q \equiv \frac{2 (1 - q_1^{3/2} - q_2^{3/2} )}{3 q_1 q_2}$ and $ q_1 \equiv 1-q_2 \equiv {m_1 \over m_1+m_2}$. This holds completely rigorously if the interpolating function is of `Bekenstein's form' (see, e.g., Zhao \& Famaey 2010; Famaey \& McGaugh 2012): for an orbit mostly in the deep-MOND ($g < a_0$) regime, the exact form of the interpolating function is of little importance, though.

Considering the expansion of the background universe, $a(t)$, the equation of motion for the relative separation (Zhao et al. 2010) is then given by:
\begin{equation}
 {d^2 \over dt^2} {\bf r}_{12}  =  K {\bf r}_{12} - \frac{m_1+m_2}{m_1} \left[\frac{{\bf F}_{12}}{m_2} \right], ~ K \equiv   \frac{{\rm d}^2a}{a {\rm d}t^2},
\end{equation}
where $F_{12}$ is their mutual force, and their distance in proper coordinates $r_{12} = |{\bf r}_1 -{\bf r}_2| = |{\bf x}_1 -{\bf x}_2| a(t)$. Note that there would be a frictional term  $({\rm d} a/ {\rm d}t) ({\rm d}{\mathbf x}_i /{\rm d}t)$ for the equation of motion in comoving coordinates.  This term does not exist when the equation is written for the proper coordinates. The remaining cosmological term is $K=n(n-1) t^{-2}$ if the cosmic expansion factor is approximated as a power-law $a(t) =(t/t_0)^{n}$. Empirically adopting an expansion of the Universe from standard cosmology, ${\rm d}a / (a {\rm d}t) = (1/14 {\rm Gyr}) \sqrt{0.667+0.333 a^{-3}}$, we find the following approximation:
\begin{equation} 
K = \frac{{\rm d}^2a}{a {\rm d}t^2} = \frac{2 }{ 3 (14 {\rm Gyr}) ^2} -\frac{2}{9 t^2} ,
\end{equation}
such that the universe was decelerating at early times when $a(t) \propto t^{2/3}$, and is exponentially accelerating at late times. Using Eqs.~1-3, orbits can then be integrated backwards in 3D as per Zhao et al.~(2010). In Sect.~3.2, we also consider the effect of the external field acting on the LG, estimated from arguments based on the local galactic escape speed from the solar neighbourhood and estimations of the actual gravity of Large Scale Structure (Famaey et al. 2007; Wu et al. 2008).

\section{Results}

\subsection{M31-Milky Way orbit}

We applied Eqs.~1-3 to the M31-MW system. Contrary to the Newtonian DM case, there is not much freedom on the masses $m_1$ and $m_2$ of the galaxies, which we parametrize in MOND through their asymptotically flat circular velocities ${V_\infty}_i = (G m_i a_0)^{1/4}$ adopted from Wu et al. (2008) and Carignan et al. (2006), see Table~1. These parameters control the period of the orbit.  The current distance of M31 has been adopted to be 770~kpc (Karachentsev et al. 2004), which influences the period too.

The other fundamental parameters are the current radial and tangential velocity of M31 w.r.t. the MW. The former is known to be $V_r = 109.3 \pm 4.4 \,$km/s, which we adopt here neglecting the small error, while the second is $V_T = 17 \pm 17 \,$km/s at 1~sigma (Sohn et al. 2012; van der Marel et al. 2012a), which we vary as stated in Table~1. This latter parameter controls the angular momentum, hence the closest approach distance. 

First, a total of 12 models have been considered (listed in Table~1), varying the parameters within realistic error bars, and the very robust conclusion is that MOND {\it unavoidably} implies that the MW and M31 had a past encounter. The evolution of the MW-M31 distance with time for the nominal parameters (1st line of Table 1) is illustrated as a dotted line on Fig.~1 (with an encounter $\sim$7 Gyr ago). 

\begin{table}   
\centering   
\begin{scriptsize}   
\centering   
\caption[]{Parameters adopted for 12 models of the M31-MW system. The first two columns give the asymptotic flat circular velocities of the MW and M31 in km/s, related to the baryonic masses through $m_b=V_\infty^4/(G a_0)$, given after in units of $10^{11} M_\odot$. The 3rd and 4th columns give the adopted  range for the current tangential velocity of M31 w.r.t. us, in km/s, and the corresponding impact parameter $b$ in kpc; the 5th and 6th columns give the look-back time of the encounter $T_{\rm enc}$, in Gyr, assuming the LG is in a negligible external field and in a non-negligible external field of 3\% of $a_0$, respectively.}   
\vspace{0.3cm}   
\label{tab-cdf}   
\begin{tabular} {l l c c c c}    
\hline     
\hline     
$V_\infty^{MW}$ ($m_b$) &  $V_\infty^{M31}$ ($m_b$) & $V_T^{M31}$   & $b$ & $T_{\rm enc}$ (no EF) & $T_{\rm enc}$ (EF$=0.03 a_0$) \\   
\hline     
             &              &                       &           &     &  \\
{\bf 180 (0.7)}      &  {\bf 225 (1.6)}       & {\bf 17-34}            & {\bf 22-48}         & {\bf -7.}    & {\bf -10.2} \\   
180 (0.7)      &  215  (1.4)      & 17-34           & 19-56          & -7.3    & -11.2 \\
180 (0.7)      &  205  (1.1)      & 17-34           & 22-55          & -7.7    &-12.3 \\   
190 (0.85)       &  250 (2.5)       & 17-34          & 19-47          & -6.     &-7.9  \\   
170 (0.55)      &  225 (1.6)       & 17-34           & 26-54          & -7.2  & -11.0 \\   
180 (0.7)      &  225 (1.6)       & 0-10           & 0-13          & -7.     &-10.2   \\ 
\hline    
   
\end{tabular}   
\end{scriptsize}   
\end{table}  

\begin{figure}
   \centering
   \includegraphics[width=8.2cm]{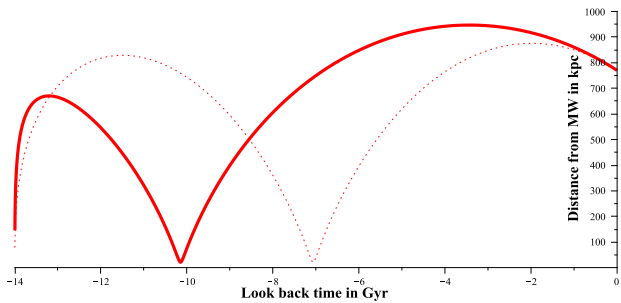}
   \caption{The figure displays the evoltion of the relative distance between the MW and M31 as a function of time for the nominal parameters in the first line of Table~1. Dotted line: negligible external field. Solid line: The external field from Large Scale Structure has been adopted to be 3\% of $a_0$ (Wu et al. 2008).}
    \end{figure}

In principle, such a past close encounter is also unavoidable for untruncated isothermal halos up to scales of several hundreds of kpc, but it would actually be {\it impossible} in the Newtonian dark halo context. Indeed, a close encounter (with a closest approach between 0 and 56~kpc) would inevitably lead to the merging of the two extended colliding halos, due to dynamical friction between their massive particles. In the context of MOND, however, the dynamical friction, while slightly enhanced within the stellar disks of the galaxies, would be negligible outside of it. As shown by the simulations of Tiret \& Combes (2008), mergers would last longer in MOND and imply multiple passages of the merging galaxies. A closest approach of $\sim$20 to 55~kpc would certainly have dynamical influences on the two colliding galaxies but would not trigger a merger in MOND, keeping in mind that stellar disks would be 60\% to 70\% of their current size at that epoch (e.g., van Dokkum et al. 2013). Note that low tangential velocities ($\sim 10$~km/s and below) in Table~1 however lead to close encounters which would probably trigger a merger in MOND too, through stellar dynamical friction. MOND thus predicts that the tangential velocity of M31 cannot be much smaller than the measured 17~km/s.

\subsection{External field effect}

In Milgromian dynamics, an important aspect is that the internal dynamics of a system is affected by the external gravitational field in which it is embedded. In the case of the MW-M31 pair, the external gravitational field from Large Scale Structures does not dominate over the internal gravity of the system along most of the orbit as long as it remains below $\sim 2$\% of $a_0$. From arguments based on the local galactic escape speed from the solar neighbourhood and estimations of the actual gravity of Large Scale Structure, the external field acting on the LG is between 1\% and 3\% of $a_0$ (Famaey et al. 2007; Wu et al. 2008). We thus recomputed the orbits for the latter value of the external field, in such a way that, when $g<0.03 a_0$, the gravity becomes essentially Newtonian again but with a renormalized $\tilde{G} \approx G/0.03$. The effect is similar to truncating a Newtonian isothermal halo (Wu et al. 2008).

To ensure a general and smooth interpolation of the two-body force between the strong, weak, and external-field dominated regimes, we considered the following empirical equation:
\begin{equation}
F_{\rm 12} \approx \frac{\tilde{G} m_1 m_2}{r_{12}^2}, ~\tilde{G} \equiv G\left[1+\left(y+\frac{g_{\rm ext}^2}{a_0^2}\right)^{-\alpha}\right]^{1 \over 2\alpha}, 
\end{equation}
where $y$ is defined as in Eq.~1. Here the parameter $\alpha$ plays the role of the interpolating function: setting $\alpha=1/2$ and $g_{\rm ext}=0$ gives back precisely Eq.~1. We checked that by adopting $\alpha=1$ and $g_{\rm ext}=0$, the closest approach and encounter epoch of M31 were almost unchanged, confirming that the exact form of the interpolating function is of little importance when most of the orbit is in the deep-MOND regime. We then considered a solution for $\alpha=1$ and $g_{\rm ext}=0.03 a_0$. For this solution, gravity thus becomes weaker in the outskirts of the orbit than without external field, and this makes the period longer. The encounter is thus pushed back to earlier times, as listed in the last column of Table~1. Interestingly, the closest approach is almost unchanged (by less than 2 kpc) compared to the no-external field case. Also, the corresponding launch speeds caluclated by integrating backwards are smaller for this non-negligible external field solution, meaning that the first encounter happens quicker after the Big Bang than in the case without external field.

For the external-field solution with the nominal values of the parameters (1st line of Table~1), the encounter took place 10.2~Gyr ago. It is compared (solid line) to the no-external field solution for the same parameters (dotted line) in Fig.~1. This is particularly interesting since this timing of the encounter roughly corresponds to the age of most dwarf satellites of the LG, bringing with it the possibility that these were created in the event as tidal dwarf galaxies (see, e.g., Kroupa et al. 2010, Kroupa 2012, Pawlowski et al. 2012, and references therein). As an example, we hereafter investigate the orbits of the Magellanic clouds in MOND, to see whether their orbits would possibly be close to pericenter at the time of the encounter.

\subsection{Magellanic clouds}

Applying the same method, in the simplifying case of a two-body configuration where the orbit of the Large Magellanic Cloud (LMC) around the MW is unperturbed by the rest of the LG, we found that  the nominal proper motion (Kallivayalil et al. 2013) of the LMC (presently at a distance of 49~kpc) would imply a period of about $\sim 3.5$~Gyr in MOND without external field, and $\sim 5$~Gyr for an external field of $0.03 a_0$ (and $\alpha=1$). In both cases, the nominal proper motions from Kallivayalil et al. (2013) neatly puts it quite close to pericenter at the time of the MW-M31 encounter. On Fig.~2, we plot (thin green line) the joint evolution of the distance to the MW of the LMC together with that of M31 (thick red line) for the solution with external field. We also show the planar shape of the orbits (lower panel of Fig.~2). While sensitive to the exact value of the proper motion, several solutions are found where the two orbits (of M31 and the LMC) would cross near the $Z$-axis, which would imply a strong orbital perturbation of the LMC and perhaps a gas shock.

\begin{figure}
   \centering
      \includegraphics[width=8.2cm]{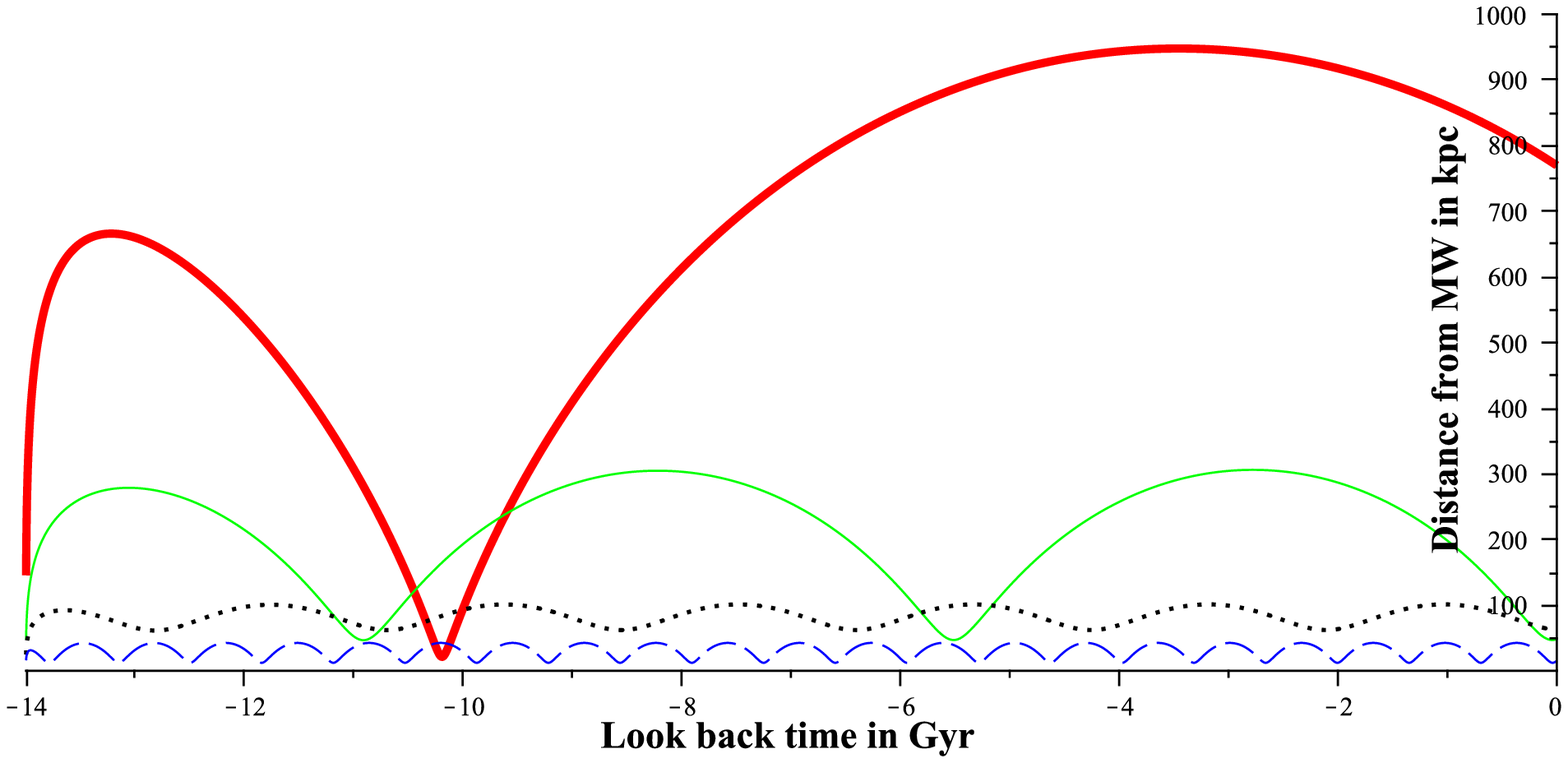}
      \includegraphics[width=10.2cm]{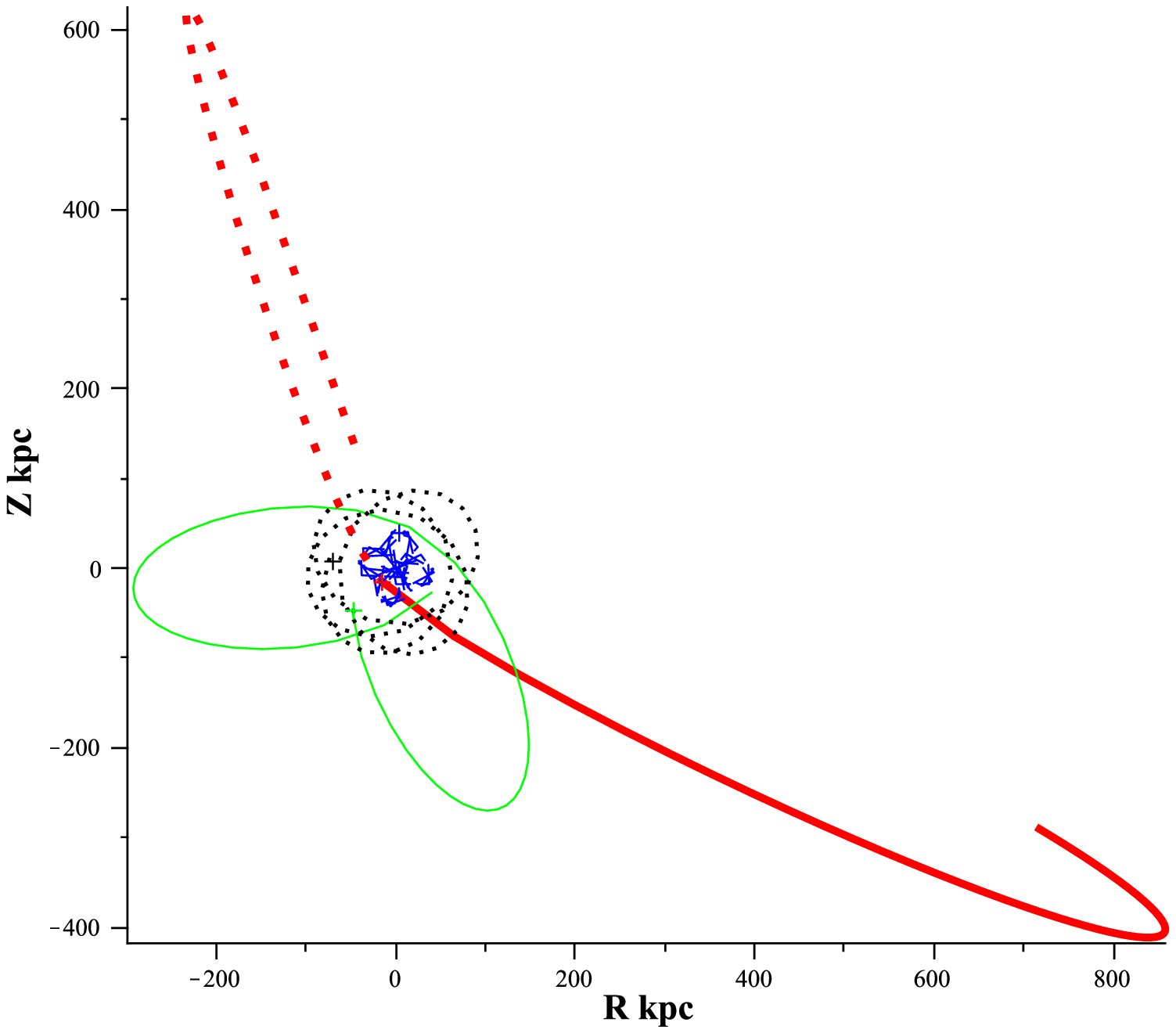}
   \caption{Upper panel: The radial distances to M31 (thick red), LMC (thin green), and SMC (dotted black) for the past 14 Gyr for the model parameters in the first line of Table~1 and an external field of $0.03 a_0$. Note the MW-M31 encounter 10.2~Gyr ago and the presence of the LMC and SMC close to pericenter at that epoch. The Sagittarius dwarf (dotted blue) orbit is also plotted for reference. Lower panel: Planar shape of the orbits w.r.t. the MW. The thick red line is the orbit of M31. The first part of the orbit is indicated as a dotted red line, and becomes a solid red line from the time of the encounter, 10.2~Gyr (marked by crosses) ago. All orbits are assumed to be on independent meridional $RZ$ planes with the horizonal axis $R=\pm \sqrt{r^2-Z^2}$ showing the offset from the MW rotation axis $Z$.}
\end{figure}

The Small Magellanic Cloud (SMC, presently at a distance of 63~kpc), considered as an independent object unperturbed by the LMC, is nearly on a circular orbit in MOND (black dotted line on Fig.~2). Note that the SMC-LMC pair is typically in a region where the external field from the MW dominates its internal dynamics, and the high relative velocity of the SMC w.r.t. the LMC then implies that the two are {\it not} bound in the MOND context. The Magellanic stream would then have to be due to recent ram-pressure stripping of gas from the SMC since the stream contain very few stars. A recent encounter of the LMC and SMC might have released some gas clouds from the gas-rich SMC, and these gas clouds were further stretched by ram-pressure with the denser clouds leading the less dense ones. In any case, at the time of the MW-M31 encounter, the three objects were thus possibly all in a region of $\sim 100$~kpc of radius, meaning that tidal interaction among these three objects (and other satellites within 100~kpc of the MW and M31) would be inevitable, and possibly the Magellanic clouds and the `disks of satellites' in the MW and M31, could be debris objects created in this event.

\section{Conclusion and discussion}

Here we have shown that if Milgromian dynamics is a valid effective description of gravity on the scales of hundreds of kpc, then the latest HST measurement of the proper motion of M31 implies that it necessarily had a past close encounter with the MW 6 to 12 Gyr ago (most probably 7 to 11 Gyr ago: see Table 1, where encounter epochs outside of this range correspond to rather unrealistic baryonic masses of M31). The closest approach would be less than $\sim$55~kpc in all cases.

Such galaxy interactions are observed to be quite common at $z \sim 1$, and Milgromian dynamics would thus imply that we might be seeing the aftermath of such an event, for instance in the peculiar orbits of the dwarf galaxies of the LG since all satellites within 100~kpc of the MW and M31 would have been affected. In the case where the encounter happens $\sim 10$~Gyr ago, it is possible that most dwarf galaxies of the LG would be {\it created} in the event.  Such a close encounter between the MW and M31 has already been suggested by Pawlowski et al. (2012) to be possibly responsible for the creation of the vast polar structure (VPOS) of satellites around the MW. Indeed the configurations of `disks of satellites' suggest that the satellite galaxies in both the MW (Lynden-Bell 1982, Zhao 1998, Kroupa et al. 2010) and partly in M31 (Ibata et al. 2013) might have been formed in tidal arms (see also Hammer et al. 2013), and hence are pure concentrations of cooled baryons which could not harbour DM halos: that tidal dwarfs exhibit missing mass in Newtonian dynamics thus only makes sense in the MOND context (see also Gentile et al. 2007). To piece together these puzzles self-consistently, it is thus particularly satisfying that such a close interaction between the MW and M31 is actually inevitable in MOND.

The LG also helps to break the often frustratingly degenerate predictions of MOND and an isothermal DM halo truncated at $\sim 100$~kpc radius.  Importantly, the M31-MW encounter scenario is self-consistent {\it only} in the absence of a particle DM halo, to avoid dynamical friction inevitably triggering a rapid merger. The lack of dynamical friction in MOND galaxy interactions has been well demonstrated by the simulations of Tiret \& Combes (2008), the key to reduce friction being to have a slightly non-head-on encounter. However, we showed that low tangential velocities (see Table~1) would lead to small impact parameters which would probably still trigger a merger through stellar dynamical friction: so, in MOND, the tangential velocity of M31 cannot really be much smaller than the measured 17~km/s. Note however that the first part of the orbit before the encounter and the actual original impact parameter are not necessarily well determined by the present modelling, since the `conserved' quantities we see today (energy and angular momentum) are only valid for point masses. Actually, there may have been a transfer of energy and angular momentum between the orbit and the internal motions during the fly-by. A small fraction of gas might also have been launched into the LG by ram pressure, which might lead to various stellar (sub)structures in the LG and in the stellar halos of both galaxies. We finally note that the origin of the thick disk of the MW is also still uncertain, and we point out that, in the MOND context, it could be the result of this same violent perturbation by M31 about $\sim 10$~Gyr ago.

\end{document}